\documentstyle[aps,eqsecnum,12pt]{revtex}
\begin{document}
\title{Theory of shot noise in space-charge limited\\ diffusive
conduction regime
}
\author{V. L. Gurevich and M. I. Muradov}
\address{Solid State Physics Department, A.F.Ioffe Institute,
194021 Saint Petersburg, Russia}
\date{\today}
\maketitle
\begin{abstract}
\baselineskip=2.5ex

As is well known, the fluctuations from a stable stationary
nonequilibrium state are described by a linearized
nonhomogeneous Boltzmann-Langevin equation. The stationary state itself
may be described by a nonlinear Boltzmann equation. The ways of
its linearization sometimes seem to be not unique. We argue that
there is actually a unique way to obtain a linear equation for
the fluctuations. In the present paper we treat as an example an
analytical theory of nonequilibrium shot noise in a diffusive
conductor under the space charge limited regime. Our approach
is compared with that of Schomerus, Mishchenko and Beenakker
[Phys.  Rev. B {\bf 60}, 5839 (1999)].  We find some difference
between the present theory and the approach of their paper and
discuss a possible origin of the difference. We believe that it
is related to the fundamentals of the theory of fluctuation
phenomena in a nonequilibrium electron gas.

\end{abstract}

\pacs{72.70.+m, 73.50.Td, 72.10.-d}
\baselineskip=3.5ex
\section{Introduction}
The present paper is devoted to the theory of shot noise in
space-charge limited diffusive conduction regime. The motivation
of the present paper can be formulated as follows. It is well
known that the fluctuations from a stable stationary
nonequilibrium state are described by a linearized
nonhomogeneous Boltzmann-Langevin equation (see, for
instance,~\cite{LL10,GK,KS0,GGK1,KS,GGK2,GGK3}). At the same
time, the stationary state itself is described by a nonlinear
Boltzmann equation. There are instances where the ways of
linearization of the nonlinear Boltzmann equation seem to be not
unique. We believe, however, that in each such case there is a
unique way to obtain the linearized Boltzmann equation for the
fluctuations and we give general considerations to find out this
way and indicate it for the particular case treated in the
present paper.

We work out a theory of nonequilibrium shot noise in a
nondegenerate diffusive conductor under space charge limited
regime. This regime is extensively discussed in the literature
(see, for instance, Refs.~\cite{LM,MG}). The current noise under
such a regime has been studied recently by Monte Carlo
simulation by Gonz$\acute{\rm a}$lez {\em et al.}~\cite{GL}.
Quite recently the noise under the same conditions has been
studied analytically by Schomerus, Mishchenko and
Beenakker~\cite{SMB}.  Their general finding was that due to the
Coulomb correlation between the electrons the shot noise is
reduced below the classical Poisson value. Both the authors of
Ref.~\cite{GL} and of Ref.~\cite{SMB} came to the conclusion
that under certain conditions the suppression factor in the
nondegenerate 3D case can be close to $1/3.$

Later on Nagaev~\cite{NAG} has shown on a special example that unlike
the $1/3$ noise reduction in degenerate systems, the noise
suppression by the Coulomb interaction is nonuniversal in
nondegenerate systems. The noise suppression in such systems may
depend on the details of electron scattering.

We agree with the conclusion~\cite{GL,SMB} that the reduction of
the shot noise power in nondegenerate diffusive conductors can
be sometimes close to the value $1/3$ predicted theoretically
for a three-dimensional (3D) degenerate electron gas.  As we
have mentioned above, we also come to some conclusions that may
prove important for the general theory of fluctuations in
nonequilibrium systems. As is well known, the fluctuation
phenomena in nonequilibrium {\it stable} systems are described
by a {\em linearized} Boltzmann equation. We would like to use
the example analyzed in detail in the present paper to show that
one should be careful performing the linearization. In
particular, there is a difference between the analytical
procedures used in Ref.~\cite{SMB} and in the present paper for
calculation of the shot noise power. We discuss the origin of
such difference and its implications. As the point leading to
the discrepancy is very subtle it demands a rather detailed
analysis that we perform in the present paper partly repeating
the calculations of Ref.~\cite{SMB} with some modifications. Our
starting point is the Boltzmann equation that is formulated for
description of the stationary state and will be applied for
analysis of the fluctuations.

\section{Boltzmann equations}
\label{sec:KE}
We will consider the simplest model, exploited in Ref.~\cite{SMB},
for the diffusion-controlled and space charge limited transport.
As the starting point we use the Boltzmann equation in the
presence of electric field
\begin{eqnarray}\label{eq:001first}
\left({\partial \over{\partial t}}+{\cal J}_{\bf p}\right)f_{\bf p}=0,
\end{eqnarray}
\begin{eqnarray}\label{eq:001}
{\cal J}_{\bf p}f_{\bf p}\equiv\left(
{\bf v}{\partial\over{\partial {\bf r}}}+e{\bf E}
{\partial\over{\partial
{\bf p}}}+I_{\bf p}\right)f_{\bf p}.
\end{eqnarray}
Here we have introduced the collision integral $I_{\bf
p}$ describing the electron scattering:
\begin{eqnarray}\label{eq:002}
I_{\bf p}f_{\bf p}=\sum_{{\bf p}\prime}
\left(W_{{\bf p}\prime{\bf p}}f_{\bf p}-
W_{{\bf p}{\bf p}\prime}f_{{\bf p}\prime}\right),
\end{eqnarray}
(we deal with the nondegenerate statistics, so that $f_{\bf
p}\ll\,1$).

Splitting the distribution function into the even and
odd parts with respect to ${\bf p}$ one gets
$$ f_{\bf p}^{\pm}={1\over 2}\left(f_{\bf p}\pm f_{-{\bf
p}}\right).$$
We assume that the collision operator acting on the even (odd) part of
the distribution function gives an even (odd) function. This may be due
either to the central symmetry of the crystal itself and the
scatterers or to the possibility to use the Born approximation for
calculation of the scattering probability. The first split equation is
\begin{eqnarray}
\label{eq:003}
{\partial f_{\bf
p}^{-}\over{\partial t}}+ {\bf v}{\partial f_{\bf p}^{+}\over{\partial
{\bf r}}}+e{\bf E} {\partial f_{\bf p}^{+}\over{\partial {\bf p}}}
=-I_{\bf p}f_{\bf p}^{-} .
\end{eqnarray}
Being interested in relatively small frequencies of fluctuations
$\omega\tau_{\bf p}\ll\,1$ where $\tau_{\bf p}$ is the characteristic
value of $I_{\bf p}^{-1}$ we can neglect the time derivative and
express $f_{\bf p}^{-}$ as
\begin{eqnarray}
\label{eq:004}
f_{\bf p}^{-}=-{{I_{\bf p}^{-1}}}
\left({\bf v}{\partial f_{\bf p}^{+}\over{\partial {\bf r}}}+e{\bf
E}{\bf v} {\partial f_{\bf p}^{+}\over{\partial \varepsilon_{\bf
p}}}\right) .
\end{eqnarray}
Inserting this expression into the second split equation
for $f_{\bf p}^+\simeq\,f(\varepsilon,{\bf r},t)$ and averaging over
the surface of constant energy in the quasimomentum space we arrive
at
\begin{eqnarray}
\label{eq:1}
\nu(\varepsilon){\partial f\over{\partial t}}-
\left({\partial \over{\partial x_\alpha}}+eE_{\alpha}
{\partial \over{\partial
\varepsilon}}\right)\nu(\varepsilon)D_{\alpha\beta}(\varepsilon)
\left({\partial \over{\partial x_\beta}}+eE_{\beta}
{\partial \over{\partial
\varepsilon}}\right)f
=-\sum_{{\bf
p}}\delta(\varepsilon-\varepsilon_{\bf p})I_{\bf p}^{({\rm inel})}f
\end{eqnarray}
where the term on the right-hand side describes the inelastic
collisions while the density of states $\nu(\varepsilon)$ and the
diffusion tensor $D_{\alpha\beta}(\varepsilon)$ are defined as
\begin{equation}
\label{eq:2}
\nu(\varepsilon)D_{\alpha\beta}(\varepsilon)=\sum_{\bf
p}\delta(\varepsilon-\varepsilon_{\bf
p})v_{\alpha}I_{\bf p}^{-1}v_{\beta},\;\;
\nu(\varepsilon)=\sum_{\bf p}\delta(\varepsilon-\varepsilon_{\bf p}).
\end{equation}
The electric field obeys the Poisson equation
\begin{equation}\label{eq:3}
\kappa\nabla\cdot{\bf E}=4\pi\,e\left[n({\bf r},t)-n^{eq}\right],\;
n({\bf r},t)=\int_0^{\infty}d\varepsilon\,\nu(\varepsilon)
f(\varepsilon ,{\bf r},t),
\end{equation}
where $\kappa$ is the dielectric susceptibility, $n^{eq}$ is the
equilibrium concentration (equal to the concentration of donors). In
what follows we neglect it as compared with the nonequilibrium
concentration $n$.

The  part of the distribution
function contributing to the current
consists of two terms proportional to the spatial and
energy derivatives of $f(\varepsilon ,{\bf r},t)$ respectively
\begin{equation}
\label{eq:4}
j_{\alpha}=e\sum_{\bf p}{\bf v}f_{\bf p}^{-}=
-e\nu(\varepsilon)D_{\alpha\beta}(\varepsilon)
\left({\partial \over{\partial x_\beta}}+eE_{\beta}
{\partial \over{\partial
\varepsilon}}\right)f.
\end{equation}
Let us consider the case $D\tau_{\varepsilon}\gg\,L^2$, where $L$
is the sample length,
$\tau_{\varepsilon}$ is the energy relaxation time (of the order of
$\left[I_{\bf p}^{({\rm inel})}\right]^{-1}$).
Then we can omit the term on
the right-hand side of Eq.~(\ref{eq:1}) which describes the energy
relaxation.  Under the same conditions we get the following Boltzmann
equation for the fluctuations of the distribution function (we remind
that we consider here low frequency fluctuations with
$\omega\ll\,I_{\bf p}\approx\,1/{\tau_{\bf p}}$ where $\tau_{\bf p}$
is the characteristic time of elastic collisions)
\begin{eqnarray}\label{eq:5}
\left({\partial \over{\partial x_{\alpha}}}+eE_{\alpha}
{\partial \over{\partial
\varepsilon}}\right)\delta j_{\omega}^{\alpha}
+e\delta E_{\omega}^{\alpha}
{\partial \over{\partial
\varepsilon}}j_{\alpha}
=ey_{\omega}(\varepsilon,x)
\end{eqnarray}
where
\begin{eqnarray}\label{eq:6}
\delta j_{\omega}^{\alpha}=e\sum_{\bf p}v_{\alpha}
\delta f_{\bf p}^{-}
=g_{\omega}^{\alpha}-e\nu(\varepsilon)D_{{\alpha}{\beta}}(\varepsilon)
\left(\left[{\partial\over{\partial
x_{\beta}}}+eE_{\beta}{\partial\over{\partial\varepsilon}}
\right]\delta f_{\omega}+
e\delta E_{\omega}^{\beta}{\partial\over{\partial\varepsilon}}f\right)
\end{eqnarray}
and where the source of the current fluctuations $g_{\omega}^{\alpha}$ is
related to the Langevin forces $y_{\bf p}^{\omega}$ as
\begin{eqnarray}\label{eq:7}
g_{\omega}^{\alpha}=e\sum_{{\bf
p}}\delta(\varepsilon-\varepsilon_{\bf p})v_{\alpha}
{{I_{\bf p}^{-1}}}y_{\bf p}^{\omega},
\end{eqnarray}
\begin{equation}
y_{\omega}(\varepsilon,x)=\sum_{{\bf
p}}\delta(\varepsilon-\varepsilon_{\bf p})y_{\bf p}^{\omega}=0.
\end{equation}
The last equality is a consequence of elasticity of scattering
that leads to the particle conservation within the surface of constant
energy in the quasimomentum space.

The correlation function of the
Langevin forces is well known~\cite{GGK3} \begin{eqnarray}\label{eq:7a}
<y_{\bf p}({\bf r})y_{{\bf p}^{\prime}}({\bf r}^{\prime})>_{\omega}=
({\cal J}_{\bf p}+{\cal J}_{{\bf p}^{\prime}})\delta_{{\bf r}{\bf
r}^{\prime}} \delta_{{\bf p}{\bf p}^{\prime}}f_{\bf p}.  \end{eqnarray}
Integrating Eq.(\ref{eq:5}) over $\varepsilon$ we obtain the
continuity equation
\begin{eqnarray}\label{eq:5a}
A{d \over{d x}}\int_0^{\infty}d\varepsilon\,\delta
j_{\omega}(\varepsilon,x)=
{d \over{d x}}\delta
J_{\omega}(x)=0,
\end{eqnarray}
which implies that the low frequency current fluctuations are
spatially homogeneous.
\section{Distribution function.}
Consider a semiconductor with a uniform cross section $A$
connecting two identical metallic electrodes. The sample's
length $L$ is assumed to be much bigger than the elastic
scattering length $l$ and much smaller than the inelastic one.
We use the 1D versions of the Boltzmann equations describing
evolution of the distribution function along the dc current
direction.

To obtain the stationary solution of Eq.~(\ref{eq:1}) in the
accepted approximation we rewrite it in the form
\begin{eqnarray}\label{eq:11}
\left({\partial \over{\partial x}}+eE
{\partial \over{\partial
\varepsilon}}\right)j(\varepsilon ,x)=\delta(x)j(\varepsilon).
\end{eqnarray}
Here we assume that the current density at
$x=0$,\quad$j(\varepsilon)$, does not vanish only for
$\varepsilon\,>\,0$. In the absence of tunneling $j(\varepsilon)$ at
the contact $x=0$ should have the property
\begin{equation}
j(\varepsilon)\rightarrow0\quad\mbox{for}\quad T\rightarrow0,
\label{j}
\end{equation}
$T$ being the temperature.
This condition should be valid, irrespective to whether
one has a Schottky barrier or an Ohmic contact.
The total current $J$ given by
Eq.~(\ref{eq:12a}) below should have of course the same property.

The solution of Eq.~(\ref{eq:11}) is a function of
the total energy ${\cal E}$ $$ {\cal E}=\varepsilon+U(x), $$ where
$U(x)=e\varphi(x)-e\varphi(0)$. It can be found
employing, for instance, the inverse differential operator $$
{1\over{\partial_x}}\Phi(x)=\int_0^xd\xi\,\Phi(\xi).  $$ One has
\begin{eqnarray}\label{eq:12}
j(\varepsilon,x)={1\over{\partial_x+eE(x)
\partial_\varepsilon}}\delta(x)j(\varepsilon)\nonumber
=\exp[e\varphi(x)\partial_{\varepsilon}]{1\over{\partial_x}}
\exp[-e\varphi(x)\partial_{\varepsilon}]\delta(x)j(\varepsilon)=
j({\cal E})
\end{eqnarray}
and $j(\varepsilon,x)$ has at a given $x$
nonzero values only if
$\varepsilon\,>\,-U(x),\;\left({\cal E}\,\ge\,0\right)$.
The total current through the sample is
\begin{eqnarray}\label{eq:12a}
J=A\int_0^{\infty}d\varepsilon\,j(\varepsilon,x)=
A\int_{-U(x)}^{\infty}d\varepsilon\,
j[\varepsilon+U(x)]=
A\int_0^{\infty}d{\cal E}\,j({\cal E}).
\end{eqnarray}
Now, from Eq.~(\ref{eq:4}) we get
\begin{eqnarray}\label{eq:12b}
f(\varepsilon,x)=-{1\over{\partial_x+eE(x)
\partial_\varepsilon}}{j(\varepsilon,x)\over{e\lambda(\varepsilon)}}+
f[\varepsilon+U(x)],
\end{eqnarray}
or
\begin{eqnarray}\label{eq:12c}
f(\varepsilon,x)=-j[{\cal E}]\int_0^xd\xi
{1\over{e\lambda[{\cal E}-U(\xi)]}}+
f[{\cal E}]
\end{eqnarray}
where
$\lambda(\varepsilon)\equiv\,\nu(\varepsilon)D(\varepsilon)$.
We have taken into account the boundary condition at the source.
Equation~(\ref{eq:12c}) can be rewritten as
\begin{eqnarray}\label{eq:13}
f[{\cal E}-U(x),x]
=\frac{\displaystyle\left[f[{\cal E}-U(L)]
\int_0^x{d\xi /{\lambda[{\cal E}-U(\xi)]
}}
+f({\cal E})
\int_x^L{d\xi /{\lambda[{\cal E}-U(\xi)]}}\right]}
{\displaystyle\int_0^L{d\xi/{\lambda[{\cal E}-U(\xi)]}}}
\end{eqnarray}
where $j(\varepsilon)$ is expressed through the difference of the
distribution function at $x=0$ and $x=L$
\begin{eqnarray}\label{eq:13add}
j({\cal E})\int_0^L{dx\over{e\lambda[{\cal E}-U(x)]}}
=f({\cal E})-f[{\cal E}-U(L)].
\end{eqnarray}

An advantage of the form we have chosen for Eq.~(\ref{eq:13}) is its
physical transparency.  The first term on the right-hand side gives the
contribution from the right boundary while the second one gives the
contribution from the left boundary.
The solution clearly demonstrates that the thermally exited carriers
injected from the contact at $x=L$ make negligible contribution
to the distribution function $f[{\cal E}-U(x),x]$
since $f({\cal E})\,\gg\,f[{\cal E}-U(L)]$ $\left({\cal
E}\,\ge\,0\right)$, for
the parameter $|U(L)|/k_{\rm B}T$ is assumed to be large.
Neglecting this term in our solution of Eq.~(\ref{eq:13}) we come to
the solution already obtained in~\cite{SMB} by assuming
absorbing boundary conditions at the current drain.

\section{Field distribution}\label{FD}
We use the Poisson equation to determine the selfconsistent
electric field that can be expressed through the obtained
distribution function. We consider such values of $x$ that
$x\,>\,x_{\overline{\varepsilon}}$, where
$$-U(x_{\overline{\varepsilon}})\,\gg\,\overline{{\cal
E}}\sim\,k_{\rm B}T$$
\begin{eqnarray}
\label{eq:8first}
-{\kappa\over{4\pi e^2}}{d^2U\over{dx^2}}&=& \int_0^{\infty}d{\cal E}\,
\nu[{\cal E}-U(x)] \nonumber\\ \times f[{\cal E}-U(x),x]
&=&\int_0^{\infty}d{\cal E}\, \nu[{\cal E}-U(x)]
j({\cal E})\int_x^L{d\xi\over{e\lambda[{\cal E}-U(\xi)]}}
\nonumber\\
&\simeq&
\nu[-U(x)]
{J\over{eA}}
\int_x^L{d\xi\over{\lambda\left[-U(\xi)\right]}}
\end{eqnarray}
Finally we get
\begin{eqnarray}\label{eq:8second}
-{\kappa\over{4\pi e^2}}
{1\over{\nu[-U(x)]}}{d^2U\over{dx^2}}=
{J\over{eA}}
\int_x^L{d\xi\over{\lambda\left[-U(\xi)\right]}}
\end{eqnarray}

Let us check that for large $x$ this equation is consistent with
the requirement of a uniform total current.  Assuming
$\nu(\varepsilon)=\nu_0\varepsilon^{d/2-1}$ and
$D(\varepsilon)=D_0\varepsilon^{s+1}$ we integrate Eq.~(\ref{eq:4})
over the transverse coordinates and energy
\begin{eqnarray}\label{eq:8}
{J\over{A}}=-e{d\over{dx}}\int_0^{\infty}d\varepsilon\,
\nu(\varepsilon)D(\varepsilon)
f(\varepsilon,x)+{eD_0\kappa (d+2s)\over{16\pi}}
[-U(x)]^s{d\over{dx}}E^2(x)
\end{eqnarray}
We  integrate by parts the second term and take into account that at
$x\,>\,x_{\overline{\cal E}}$ we can neglect ${\cal E}$
as compared to $|U(x)|$ and use the Poisson equation
(\ref{eq:3}). The first term in Eq.(\ref{eq:8})
can be simplified in the same way [note that due to
Eq.~(\ref{eq:12c}) the distribution function $f(\varepsilon, x)$
has nonzero values only for $\varepsilon\,>\,-U(x)$ ]
\begin{eqnarray}\label{eq:9}
\int_{-U(x)}^{\infty}d\varepsilon\,\nu(\varepsilon)D(\varepsilon)
f(\varepsilon,x)=
\int_0^{\infty}d{\cal E}\,\nu[{\cal E}-U(x)]
D[{\cal E}-U(x)]
f({\cal E}-U(x),x)\nonumber\\
=D[-U(x)]\int_0^{\infty}d{\cal E}\,\nu[{\cal E}-U(x)]
f({\cal E}-U(x),x)=D[-U(x)]{\kappa\over{4\pi e}}{d\over{dx}}E
\end{eqnarray}
Here in the second equality we have taken into account  that
${\cal E}\,\ll\,|U(x)|$.  Inserting
Eq.~(\ref{eq:9}) into Eq.~(\ref{eq:8}) we get the following simplified
equation
\begin{eqnarray}\label{eq:9a}
{4\pi|J|\over{D_0\kappa
A}}={d\over{dx}}\left([-U]^{s+1}{dE\over{dx}}\right)+
|e|{2s+d\over{4}}[-U]^s{dE^2\over{dx}}
\end{eqnarray}
It can be used to verify the self-consistency of our approach.
Indeed, multiplying Eq.~(\ref{eq:8second}) by $U^{s+d/2}$
and taking the derivative we arrive at
Eq.~(\ref{eq:9a}) that has been obtained from the equation
for the current.
A dimensionless version of Eq.~(\ref{eq:9a}) is
\begin{equation}\label{dimless0}
\chi^s\left({d-2\over{2}}\chi^\prime\chi^{\prime\prime}-
\chi\chi^{\prime\prime\prime}\right)=1
\end{equation}
where the dimensionless potential $\chi$ is related to $\varphi$ by
\begin{equation}\label{dimless1}
\varphi=\left(4\pi|J|L^3\over{D_0\kappa A|e|^{s+1}}\right)^{1/(s+2)}\chi(x/L).
\end{equation}
\section{Current and field fluctuations}
In what follows we consider the particular cases $s=0,\quad$
$D(\varepsilon)=D_0\varepsilon$;
$s=-1/2,$ $D(\varepsilon)=D_0\varepsilon^{1/2}$; and
$s=1/2,$ $D(\varepsilon)=D_0\varepsilon^{3/2}$.
We begin with investigation of the energy-independent-%
scattering-time case
$s=0$. This case can be related to the scattering of electrons by
the neutral impurities, such as hydrogen-like shallow donor and
acceptor states. The scattering is
analogous to the scattering of electron by a hydrogen atom~\cite{LanLif}
(with the effective Bohr radius $a_B$). The scattering
cross-section turns out to be about $2\pi\hbar/(pa_B)$ times larger
than the geometrical cross-section $\pi a_B^2$ (that would result
in an energy-independent scattering time).

In the case of defects with deep energy levels we encounter a
short-range scattering
potential with the scattering length about atomic length.
The scattering
cross-section does not depend on the energy. As a result, the
scattering rate is proportional to the electron density of
states $\varepsilon^{1/2}$ and the diffusion coefficient
$v^2\tau$ is
proportional to $\varepsilon^{1/2}$, i.e. $s=-1/2$. (This is
one of the main scattering mechanisms in
metals since the scattering length there is determined by the radius of
screening which is of the order of interatomic distance.)
The case $s=-1/2$ (that, in particular, describes elastic
scattering by acoustic phonons) and $s=1/2$
will be discussed at the end of this
section.
\subsection{Energy independent scattering time}
Integrating
Eq.~(\ref{eq:6}) over $\varepsilon$, we get
\begin{eqnarray}\label{eq:14}
{1\over A}\left(\delta J_{\omega}-G_{\omega}\right)=-
e{d\over{dx}}\int_{-U(x)}^{\infty}d\varepsilon\,\nu(\varepsilon)
D(\varepsilon)\delta
f_{\omega}(\varepsilon,x)+{eD_0d\kappa\over{8\pi}}{d\over{dx}}E(x)\delta E_{\omega}(x)
\end{eqnarray}
Note that the Fourier transform of
the current fluctuations $\delta J_{\omega}$ due to Eq.(\ref{eq:5}) is
spatially homogeneous.
Here $G_{\omega}$ is the current fluctuations source integrated
over the energy and transverse coordinates
\begin{eqnarray}\label{eq:14a}
G_{\omega}(x)=\int_0^{\infty}d\varepsilon\,d{{\bf r}_{\perp}}
g_{\omega}(\varepsilon,{\bf r})
\end{eqnarray}
\begin{eqnarray}\label{eq:14b}
<G(x)G(x^{\prime})>_{\omega}=
e^2\int_0^{\infty}d\varepsilon\int_0^{\infty}d\varepsilon^{\prime}
\sum_{{\bf p}{\bf p}^{\prime}}
\delta(\varepsilon-\varepsilon_{\bf p})
\delta(\varepsilon^{\prime}-\varepsilon_{{\bf p}^{\prime}})
{v_x}{v_x}^{\prime}\nonumber\\
\times{1\over{I_{\bf p}}}
{1\over{I_{{\bf p}^{\prime}}}}\int\,
d{\bf r}_{\perp} d{\bf r}\prime_{\perp}
<y_{\bf p}y_{{\bf p}^{\prime}}>_{\omega}
\end{eqnarray}
The odd (with respect to ${\bf p}\rightarrow-{\bf p}$) part of the
distribution function vanishes after one inserts it into the
correlation function (\ref{eq:7a}) of the Langevin forces and
subsequent integration over $\bf p$ and ${\bf p}^{\prime}$.  As a
result, we are left with the integral of the even function
\begin{eqnarray}
\label{eq:14dfirst}
<G(x)G(x^{\prime})>_{\omega}=\delta_{xx^{\prime}}<G^2(x)>_{\omega},
\end{eqnarray}
\begin{eqnarray}\label{eq:14dsecond}
\langle G^2(x)\rangle_{\omega}=2e^2A\int_0^{\infty}d\varepsilon
f(\varepsilon,x)\sum_{\bf p}
\delta(\varepsilon-\varepsilon_{\bf p})
{v_x}
{1\over{I_{\bf
p}}}{v_x}
=2e^2A\int_{0}^{\infty}d\varepsilon\,
\nu(\varepsilon)D(\varepsilon)f(\varepsilon,x) .
\end{eqnarray}
The second term on the right hand side of Eq.~(\ref{eq:14}) can be
simplified in the same way as Eq.~(\ref{eq:9})
\begin{eqnarray}\label{eq:15}
\int_0^{\infty}d\varepsilon\,\nu(\varepsilon)
D(\varepsilon)\delta
f_{\omega}(\varepsilon, x)=D(-U(x)){\kappa\over{4\pi e}}
{d\over{dx}}\delta E_{\omega}
\end{eqnarray}
and finally we get the equation for $\delta E_{\omega}$
\begin{eqnarray}\label{eq:16}
{d\over{dx}}\left(U(x){d\over{dx}}\delta
E_{\omega}(x)\right) +e{d\over{2}}{d\over{dx}}E(x)\delta
E_{\omega}(x)= {4\pi\over{AD_0\kappa}}(\delta J_{\omega}-G_{\omega}).
\end{eqnarray}
In order to justify the simplification in Eq.~(\ref{eq:15})
we will show that $\delta f_{\omega}(\varepsilon, x)$ is also a
function acquiring nonzero values only at
$\varepsilon\,>\,-U(x)$. Indeed, from Eq.~(\ref{eq:5}) and
Eq.~(\ref{eq:6}) one can get the following solutions
\begin{eqnarray}\label{eq:16for_delta_j}
\delta
j_{\omega}[\varepsilon-U(x),x]=\delta U_{\omega}(x)
{\partial\over{\partial\varepsilon}}j(\varepsilon)+\Delta j(\varepsilon)_{\omega},
\end{eqnarray}
\begin{eqnarray}\label{eq:16for_delta_f}
\delta
f_{\omega}[\varepsilon-U(x),x]=\int_x^Ld\xi\,
e\delta E_{\omega}(\xi){\partial\over{\partial\varepsilon}}
f[\varepsilon-U(\xi),\xi]\nonumber \\
-\int_x^Ld\xi\,
{g_{\omega}[\varepsilon-U(\xi),\xi]-\delta
j_{\omega}[\varepsilon-U(\xi),\xi]
\over{e\lambda[\varepsilon-U(\xi)]}}
\end{eqnarray}
which show that $\delta f$ has the above mentioned
property. Here $\Delta j(\varepsilon)$ are the fluctuations of the
current at the left boundary $x=0$. The fluctuations of the
distribution function $\Delta f(\varepsilon)$ at the right boundary
are assumed to be zero. If we assume $\lambda(\varepsilon)$ to be a
constant (not depending on the energy), taking into account
Eqs.~(\ref{eq:16for_delta_j}) and (\ref{eq:16for_delta_f}) and the
equation $\delta f_{\omega}(\varepsilon,0)=0$ we immediately arrive
at the result obtained by Nagaev~\cite{NAG}
\begin{equation}
\Delta J={1\over{L}}\int_0^L\,dx\int\,d\varepsilon
g[\varepsilon-U(x),x] .
\end{equation}

\subsection{Comparison with approach of Ref.~[4]}\label{CB}
Now we embark on setting forth the crucial point of the paper.
Eq.~(\ref{eq:16}) {\it does not coincide} with the equation for the
field fluctuations obtained in~\cite{SMB} by direct linearization
of Eq.~(\ref{eq:9a}) for $s=0$
\begin{eqnarray}
\label{eq:9lin}
{d\over{dx}}\left[\delta U_{\omega}(x){d\over{dx}}
E(x)\right]
+{d\over{dx}}\left[U(x){d\over{dx}}\delta
E_{\omega}(x)\right] \nonumber \\
+e{d\over{2}}{d\over{dx}}E(x)\delta
E_{\omega}(x)
={4\pi\over{AD_0\kappa}}(\delta J_{\omega}-G_{\omega}) .
\end{eqnarray}
It is necessary to understand the origin of this discrepancy.

First, we adopt for the time being the scheme of Ref~\cite{SMB} and
reconsider Eq.~(\ref{eq:8}) for the current
\begin{eqnarray}\label{dob1}
{J\over{A}}=
-e{d\over{dx}}\int_{-U(x)}^{\infty} d\varepsilon\,
\nu(\varepsilon)D(\varepsilon)
f(\varepsilon,x)\nonumber\\
+{3\over 2}D_0e^2E(x)\int_{-U(x)}^{\infty}
d\varepsilon \nu(\varepsilon)f(\varepsilon,x) .
\end{eqnarray}
For the total current (the d.c. current plus fluctuations) the equation
reads \begin{eqnarray} \label{dob2} {J+\delta J-G\over{A}}=
-e{d\over{dx}}\int_{-U(x)-\delta U(x)}^{\infty} d\varepsilon\,
\nu(\varepsilon)D(\varepsilon)
[f(\varepsilon,x)+\delta f(\varepsilon,x)]\nonumber\\
+{3\over 2}D_0e^2[E(x)+\delta E(x)]\int_{-U(x)-\delta U(x)}^{\infty}
d\varepsilon \nu(\varepsilon)[f(\varepsilon,x)+\delta f(\varepsilon,x)]
\end{eqnarray}
Taking into account Eq.~(\ref{dob1}) we get the following linearized
equation
\begin{eqnarray}
\label{dob3}
{\delta J-G\over A}=-e{d\over{dx}}
\int_{-U(x)}^{\infty} d\varepsilon\,
\nu(\varepsilon)D(\varepsilon)
\delta f(\varepsilon,x)\nonumber\\
+{3\over 2}D_0e^2E(x)
\left\{\int_{-U(x)}^{\infty}
d\varepsilon \nu(\varepsilon)\delta f(\varepsilon,x)
+\delta U {\delta\over{\delta U(x)}}\int_{-U(x)}^{\infty}
d\varepsilon \nu(\varepsilon)f(\varepsilon,x)\right\}\nonumber\\+
{3\over 2}D_0e^2\delta E(x)
\int_{-U(x)}^{\infty}
d\varepsilon \nu(\varepsilon)f(\varepsilon,x)
-e{d\over{dx}}\delta U {\delta\over
{\delta U(x)}}\int_{-U(x)}^{\infty} d\varepsilon\,
\nu(\varepsilon)D(\varepsilon)
f(\varepsilon,x).
\end{eqnarray}
If one linearized the Poisson equation in the spirit of
Ref.~\cite{SMB} one would see that the term in the curly brackets
in Eq.~(\ref{dob3}) would coincide with $(\kappa/4\pi e)(d\delta
E/dx)$, so that \begin{eqnarray} \label{dob4} {\kappa\over{4\pi
e}}{d\delta E\over{dx}}=\int_{-U(x)}^{\infty} d\varepsilon
\nu(\varepsilon)\delta f(\varepsilon,x) +\delta U {\delta\over{\delta
U(x)}}\int_{-U(x)}^{\infty} d\varepsilon
\nu(\varepsilon)f(\varepsilon,x).
\end{eqnarray}
Simplifying the first, third and fourth
terms on the right-hand side of Eq.~(\ref{dob3}) with the help of
Eq.~(\ref{eq:15}) and inserting instead of the term in the curly
brackets $(\kappa/4\pi e)(d\delta E/dx)$  we arrive at
\begin{equation}
{\delta J-G\over A}=
e{d\over{dx}}\left(D_0U(x){\kappa\over{4\pi}}{d\delta
E\over{dx}}\right)+{3\over 2}D_0e{\kappa\over{4\pi}}{d\over{dx}}E\delta
E+eD_0{\kappa\over{4\pi}}{d\over{dx}}\delta
U{\delta\over{\delta U}}\left[U{dE\over{dx}}\right]
\end{equation}
One
can see that the last term on the right-hand side of this equation
coincides with the first term on the left-hand side of
Eq.~(\ref{eq:9lin}). To avoid confusion note that we believe
Eq.~(\ref{dob4}) to be also wrong.  We have written it here only for
the sake of detailed comparison with the approach of
Ref.~\cite{SMB}.  We believe that the correct Poisson equation for
the fluctuation field is
\begin{eqnarray} {\kappa\over{4\pi e}}{d\delta
E\over{dx}}=\int_{-U(x)}^{\infty} d\varepsilon \nu(\varepsilon)\delta
f(\varepsilon,x).
\end{eqnarray}

Now we will add in Eq.~(\ref{eq:8}) for the d.c. current the
terms that actually vanish as they are proportional to the
integrals of the distribution function over $\varepsilon$ with the
upper limit $-U(x)$ whereas the distribution function
$f(\varepsilon,x)=0$ for $\varepsilon<-U(x)$.  The point is that when
we calculate the fluctuations by replacement $U(x)\rightarrow U(x)
+\delta U(x)$ {\em they will give a nonvanishing result}. We have
\begin{eqnarray}
\label{dob5}
{J\over{A}}=-e{d\over{dx}}\int_0^{-U(x)} d\varepsilon\,
\nu(\varepsilon)D(\varepsilon)
f(\varepsilon,x)-e{d\over{dx}}\int_{-U(x)}^{\infty} d\varepsilon\,
\nu(\varepsilon)D(\varepsilon)
f(\varepsilon,x)\nonumber\\
+{3\over 2}D_0e^2E(x)\int_{0}^{-U(x)}
d\varepsilon \nu(\varepsilon)f(\varepsilon,x)+
{3\over 2}D_0e^2E(x)\int_{-U(x)}^{\infty}
d\varepsilon \nu(\varepsilon)f(\varepsilon,x) .
\end{eqnarray}
Rewriting this equation for the total current we get
\begin{eqnarray}
\label{dob6}
{J+\delta J-G\over{A}}=-e{d\over{dx}}\int_0^{-U(x)-\delta U} d\varepsilon\,
\nu(\varepsilon)D(\varepsilon)
[f(\varepsilon,x)+\delta f(\varepsilon,x)]\nonumber\\
-e{d\over{dx}}\int_{-U(x)-\delta U(x)}^{\infty} d\varepsilon\,
\nu(\varepsilon)D(\varepsilon)
[f(\varepsilon,x)+\delta f(\varepsilon,x)]\nonumber\\
+{3\over 2}D_0e^2[E(x)+\delta E(x)]\int_{0}^{-U(x)-\delta U(x)}
d\varepsilon \nu(\varepsilon)[f(\varepsilon,x)+\delta
f(\varepsilon,x)]\nonumber\\
+{3\over 2}D_0e^2[E(x)+\delta E(x)]\int_{-U(x)-\delta U(x)}^{\infty}
d\varepsilon \nu(\varepsilon)[f(\varepsilon,x)+\delta f(\varepsilon,x)].
\end{eqnarray}
Linearizing this equation and using relations like
\begin{eqnarray}
\label{dob7}
\delta U(x){\delta\over{\delta U(x)}}\int_{-U(x)}^{\infty} d\varepsilon\,
\nu(\varepsilon)D(\varepsilon)
f(\varepsilon,x)=-\int_{-U(x)}^{-U(x)-\delta U(x)} d\varepsilon\,
\nu(\varepsilon)D(\varepsilon)
f(\varepsilon,x)
\end{eqnarray}
we arrive at Eq.~(\ref{eq:14}) that has been
derived above. We see that the cancellation of the linear in $\delta U$
contributions in Eq.~(\ref{dob6}) is due to the terms that vanish in
the equation for the d.c. current but should be taken into account when
one considers fluctuations. This is why the linearization of
Eq.~(\ref{eq:9a}) leads to Eq.~(\ref{eq:9lin}) that we believe to be
wrong as it does not take into account all the sources of fluctuation
or, in other words, all the terms in Eq.~(\ref{dob5}) containing $U(x)$.

The solution of Eq.~(\ref{eq:16}) with boundary conditions
\begin{eqnarray}\label{eq:17}
E(x)\delta E_{\omega}(x)\left|_{x\rightarrow
0}\right.\rightarrow 0,\\\nonumber
U(x){d\over{dx}}\delta E_{\omega}(x)\left|_
{\raise-8pt\hbox{${x\rightarrow0}$}}\right.\rightarrow 0
\end{eqnarray}
is
\begin{eqnarray}\label{eq:18}
-{AD_0\kappa\over{4\pi}}\delta E_{\omega}(x)=
U^{d/2}(x)\left[C+
\int_0^x{d\xi\over{U^{d/2+1}(\xi)}}\int_0^{\xi}d\eta\,(\delta
J_{\omega}-G(\eta)_{\omega})\right]
\end{eqnarray}
where $C$ is an integration constant.
Requiring a nonfluctuating applied voltage
$$
\int_0^Ldx\,\delta E_{\omega}=0
$$
we get from Eq.(\ref{eq:18}) that the constant is
\begin{equation}\label{eq:1_add}
C=\int_0^Ldx\,\left({\psi(x)\over{\psi(L)}}-1\right)
{1\over{U^{d/2+1}(x)}}\int_0^{x}d\xi\,(\delta
J_{\omega}-G_{\omega}(\xi)),
\end{equation}
where
\begin{equation}\label{eq:2_add}
\psi(x)=\int_0^xd\xi\,U^{d/2}(\xi).
\end{equation}
Now we require at the right boundary
\begin{eqnarray}\label{eq:3_add}
{d\over{dx}}\delta
E_{\omega}(x)\left|_{\raise-8pt\hbox{$x=L$}}\phantom{{\tilde{E}}}\right.=0
\end{eqnarray}
and get
\begin{eqnarray}\label{eq:19}
\delta J_{\omega}={1\over{Z}}
\int_0^Ldx\,\Pi(x)G_{\omega}(x)
\end{eqnarray}
where
\begin{eqnarray}\label{eq:19_add}
Z=L+{dU^{\prime}(L)U^{d/2}(L)\over{2\psi(L)}}\int_0^Ldx\,
{x\psi(x)\over{U^{d/2+1}(x)}},
\end{eqnarray}
\begin{eqnarray}\label{eq:19_ad_2}
\Pi(x)=1+{dU^{\prime}(L)U^{d/2}(L)\over{2\psi(L)}}\int_x^Ld\xi\,
{\psi(\xi)\over{U^{d/2+1}(\xi)}}.
\end{eqnarray}
Then the noise power $P$ is
\begin{eqnarray}\label{eq:20}
P={2\over{Z^2}}
\int_0^Ldx\,\Pi^2(x)
<G^2(x)>_{\omega}.
\end{eqnarray}
Since, according to Eq.~(\ref{eq:14dsecond})
\begin{eqnarray}\label{eq:21}
<G^2(x)>_{\omega}=2e^2A\int_0^{\infty}d\varepsilon\,\nu(\varepsilon)
D(\varepsilon)f(\varepsilon,x)
=2e^2AD_0U(x){\kappa\over{4\pi}e}
{d^2U\over{dx^2}} .
\end{eqnarray}
Finally we arrive at
\begin{eqnarray}\label{eq:22}
P={4AD_0\kappa\over{4\pi Z^2}}
\int_0^Ldx\,\Pi^2(x)U(x){d^2U\over{dx^2}}.
\end{eqnarray}
The potential distribution can be found following method
of Ref.~\cite{SMB}, {\em i.e.} solving Eq.(\ref{eq:9a}) with
boundary condition  Eq.(\ref{eq:8second}) at $x=L$. Using
Eqs.~(\ref{eq:2_add}), (\ref{eq:19_add}), (\ref{eq:19_ad_2}) and
(\ref{eq:22}) we calculate the suppression factor $P/P_{\rm Poisson}$.
For physically relevant different values of the dimensionality
$d$ we get
\begin{eqnarray}\label{eq:poisson} P/P_{\rm Poisson}=
\left\{\begin{array}{lcr}
0.3188 & \mbox{for} &d=3,\\
0.4512 &\mbox{for}  &d=2,\\
0.682  &\mbox{for}  &d=1.\end{array}\right.
\end{eqnarray}
Thus, in this particular case our results differ from those
calculated in Refs.~\cite{SMB} both analytically (that, in our
opinion, is of principal importance) and numerically (although in
this particular case the difference is not great). Naturally, there
is essentially no difference with the results calculated within
an ensemble Monte Carlo scheme in Ref.~\cite{GL}.

\subsection{Energy dependent scattering time}
Here we calculate the noise power for $s=\pm\,1/2$ and
$d=3$. The equation for the fluctuations is
\begin{eqnarray}
\label{eq:sc1}
-{4\pi\over{\kappa D_0A}}(\delta J_{\omega}-G_{\omega})=
{d\over{dx}}\left[(-U)^{s+1}{d\delta
E_{\omega}\over{dx}}\right]
-e{2s+d\over 2}(-U)^s{d\over{dx}}\left(E\delta E_{\omega}\right).
\end{eqnarray}
Introducing the dimensionless potential $\chi$ given by~%
Eq.(\ref{dimless1}) and the fluctuation of the field $\Delta E$
\begin{equation}
\delta E(x)={1\over{L}} \left({4\pi|J|L^3\over{\kappa
D_0A|e|^{s+1}}}\right)^{1/(s+2)} \Delta E\left({x\over L}\right),
\end{equation}
one can rewrite Eq.~(\ref{eq:sc1}) as
\begin{equation}\label{eq:dif_fl0} \Delta
E^{\prime\prime}+\left(1-
{d\over{2}}\right){\chi^{\prime}\over{\chi}}\Delta E^{\prime}-
\left(s+
{d\over{2}}\right){\chi^{\prime\prime}\over{\chi}}\Delta E=
{1\over{\chi^{s+1}}}{(G-\delta J)\over{|J|}}.
\end{equation}
We assume $s=-1/2$, $d=3$ and get
\begin{equation}\label{eq:dif_fl}
\Delta E^{\prime\prime}-
{1\over{2}}{\chi^{\prime}\over{\chi}}\Delta E^{\prime}-
{\chi^{\prime\prime}\over{\chi}}\Delta E=\chi^{-1/2}{(G-\delta J)\over{|J|}}.
\end{equation}
This equation differs from that derived in Ref.~\cite{SMB}
while equation for the potential $\chi$ coincides with
\begin{equation}
{1\over{2\chi^{1/2}}}\chi^{\prime}\chi^{\prime\prime}-
\chi^{1/2}\chi^{\prime\prime\prime}=1.
\end{equation}
To calculate the Green's function of Eq.~(\ref{eq:dif_fl}) we
need the function $\psi_1(x)$ obeying the homogeneous equation
\begin{equation}\label{eq:psi_1}
\psi_1^{\prime\prime}-
{1\over{2}}{\chi^{\prime}\over{\chi}}\psi_1^{\prime}-
{\chi^{\prime\prime}\over{\chi}}\psi_1=0
\end{equation}
and satisfying the boundary condition
$\psi_1^{\prime}|_{x=0}=0$.
The second function $\psi_2$ obeying the boundary condition
$\psi_2^{\prime}|_{x=L}=0$ can be expressed through the
functions $\chi$ and $\psi_1$
\begin{equation}\label{eq:psi_2}
\psi_2(x)=-\psi_1\left[{\chi^{1/2}(1)\over{\psi_1(1)\psi_1^{\prime}(1)}}+
\int_x^Ld\xi\,{\chi^{1/2}(\xi)\over{\psi_1^2(\xi)}}\right].
\end{equation}
The solution of Eq.~(\ref{eq:dif_fl}) can be written using the
Green's function
\begin{equation}\label{eq:Green's_func}
G(x,x^{\prime})={1\over{\chi^{1/2}(x^{\prime})}}
\left[\theta(x-x^{\prime})\psi_1(x^{\prime})\psi_2(x)+
\theta(x^{\prime}-x)\psi_1(x)\psi_2(x^{\prime})\right]
\end{equation}
as
\begin{equation}\label{eq:field}
\Delta
E=\int_0^1dx^{\prime}\,G(x,x^{\prime})
{(G(x^{\prime})-\delta J)\over{\chi^{1/2}(x^{\prime})|J|}}.
\end{equation}
Requiring a nonfluctuating applied voltage we get
\begin{equation}\label{eq:j_fl}
\delta J={1\over{Z}}\int_0^1dx\,{G(x)\over{\chi(x)}}\Pi(x)
\end{equation}
where
\begin{equation}\label{eq:Pi}
\Pi(x)=\psi_1(x)\int_x^1d\xi\,\psi_2(\xi)+
\psi_2(x)\int_0^xd\xi\,\psi_1(\xi),
\end{equation}
\begin{equation}\label{eq:Z}
Z=\int_0^1dx\,{\Pi(x)\over{\chi(x)}}.
\end{equation}
Expressing the correlation function $<G^2(x)>$ through $\chi$ we get
for the shot noise power reduction factor
\begin{equation}\label{eq:P/P_poiss}
{P\over P_{\rm Poisson}}={2\over{Z^2}}\int_0^1dx\,
{\chi^{\prime\prime}(x)\over{\chi^{3/2}(x)}}\Pi^2(x).
\end{equation}
We determine potential $\chi$ following
Ref.~\cite{SMB} and find numerically $\psi_1$ from
Eq.~(\ref{eq:psi_1}). Now the functions $\psi_2$, $\Pi$ and the
constant $Z$ can be found from
Eqs.~(\ref{eq:psi_2}), (\ref{eq:Pi}), and (\ref{eq:Z}). The
reduction factor can be evaluated as
\begin{equation}\label{eq:red_fact}
P/P_{\rm Poisson}=0.4257
\end{equation}
which is about $10\%$ larger than the result obtained in
Ref.~\cite{SMB}. As indicated by Gonz$\acute{\rm a}$lez,
Gonz$\acute{}\!{\rm a}$lez, Mateos, Pardo, Reggiani, Bulashenko,
and Rub$\acute{\rm{\imath}}$~\cite{GL}, they obtained for
$s=-1/2$ as a result of numerical simulation
\begin{equation}\label{eq:red_fact'}
P/P_{\rm Poisson}=0.42 \mbox{ --- } 0.44.
\end{equation}
One can see that this interval is noticeably nearer to the value given
by Eq.~(\ref{eq:red_fact}) than the result of Ref.~\cite{SMB}.

In the case $s=1/2$ the reduction factor can be evaluated as
\begin{equation}\label{eq:rf_half}
P/P_{\rm Poisson}=0.1974.
\end{equation}
that is slightly smaller than the result of Ref.~\cite{GL}.
\section{Conclusion}
In summary, we have developed an analytical theory of shot noise in a
diffusive conductor under the space charge limited regime. We find
difference between the present theory and the approach
developed earlier and indicate a possible origin of the difference.

Several conclusive remarks. The calculated nonequilibrium shot noise
power in a nondegenerate diffusive semiconductor for two types of
physically relevant elastic scattering mechanisms turned out to be
very close to the ones obtained in numerical simulations by the
authors of Ref.~\cite{GL}. The computed noise suppression factor
$P/P_{\rm Poisson}$ for the case of an energy-independent scattering
time is also rather close to the analytical results obtained
earlier by Schomerus {\it et al.}~\cite{SMB}. However, for an
energy-dependent scattering the numerical difference between our
results and the ones of Ref.~\cite{SMB} is considerable.

Let us clarify once more the point as to why the authors of
Ref.~\cite{SMB} arrived at the equations that differ from ours.
As an example we take the Poisson equation. According to
Ref.~\cite{SMB} one could write
\begin{eqnarray}\label{primer}
n=\int_{-U(x)}^{\infty} d\varepsilon \nu(\varepsilon)
f(\varepsilon,x),
\end{eqnarray}
where $n,U$ are the exact {\em total} concentration and potential
energy, $f$ is the {\em total} distribution function(the mean value
{\em plus} the fluctuating part). Linearization of this equation leads
to equations of Ref.~\cite{SMB}. The authors of Ref.~\cite{SMB}                could
have argued that since the voltages in the reservoirs do not fluctuate
and $U$ is set to zero at the left boundary and since the total energy
${\cal E}=\varepsilon+U$ is and remains positive therefore the total
distribution function is zero for $\varepsilon\,<\,-U$.

Our point is that one cannot justify Eq.~(\ref{primer}) for the
total values of these variables including the stationary and
fluctuating parts. This is readily seen from the fact that the
fluctuating part of the distribution function itself depends
implicitly on the mean value of the distribution function
through the correlation function. One should bear in mind that
an equation involving both the mean and the fluctuating
quantities must be regarded symbolically. Indeed, such an
equation is in fact equivalent to two equations: one for the
mean values and the other for the fluctuating part. Regarded
literally it can lead to confusion. For instance, analyzing the
equation
$$
\overline{n}+\delta
n=\int_{-\overline{U}-\delta
U}d\varepsilon\nu(\varepsilon)(\overline{f}+\delta f)
$$
one could have come to a wrong conclusion that the mean value
$\overline{n}$ depends on such an average as $\overline{\delta
U\delta f}$.


A few words about the boundary conditions for the potential.
The boundary conditions used are not applicable within the
length $R_V=\sqrt{\kappa V/4\pi en(0)}$ near the electrodes. As the
nonequilibrium noise power is a bulk property [mark, for
instance, integration over the coordinate in Eq.~(\ref{eq:P/P_poiss})]
this approximation is justified since we assume that the sample's
length $L$ is much bigger than $R_V$.

Being interested in analysis of the fluctuation phenomena in the
simplest situation of the space-charge limited diffusive
conduction regime, we have not taken into account the
electron-electron collisions. Meanwhile, such collisions can
bring about additional electron-electron correlation~\cite{GGK3}
which one should consider treating a more general case.

\acknowledgements
The authors are grateful to K. E. Nagaev for communicating to
them his views concerning the role of boundary conditions at the
contact between metal and semiconductor for the space-charge
limited diffusive conduction.

The authors acknowledge the support for this work
by the Russian National Fund of Fundamental Research (Grant
No~00-15-96748).

\end{document}